\documentclass[aps,prb,twocolumn,showpacs,preprintnumbers]{revtex4}

\usepackage{graphicx}
\usepackage{dcolumn}
\usepackage{bm}

\usepackage{amssymb}

\begin{document}

%
%


\title{Low-Temperature Magnetic Anomaly in Magnetite}




\author{Z. \v{S}vindrych}
 \email{svindr@fzu.cz}

\author{Z. Jan\accent23 u}
 \affiliation{Institute of Physics of the AS CR, v.v.i., Na Slovance 2, 182 21 Prague 8, Czech Republic}
 

\author{A. Koz{\l}owski}
\affiliation{Faculty of Physics and Applied Computer Science, AGH-University of Science and Technology, Krak\'ow, Poland}

\author{J. M. Honig}
\affiliation{Department of Chemistry, Purdue University, 1393 Brown Building, West Lafayette, Indiana 47907-8024, USA}


\date{\today}
\begin{abstract}
We have studied experimentally the responses of high quality single crystals of stoichiometric synthetic magnetite to applied weak dc and ac magnetic fields in the range of 6~K to 60~K, far below the Verwey transition.
The results can be compared to so called Magnetic After Effects (MAE) measurements, which are the most extensive magnetic measurements of magnetite at these temperatures.
We present a novel point of view on the relaxation phenomena encountered at these temperatures - the Low Temperature Anomaly, addressing the striking difference between the results of conventional ac susceptibility measurements and those accompanying MAE measurements, i.e. periodic excitations with strong magnetic pulses. We also draw a connection between this anomaly and the so called glass-like transition and discuss possible mechanisms responsible for these effects.
\end{abstract}



\maketitle

\section{Introduction}
Unusual physical properties of magnetite (Fe$_3$O$_4$), the oldest known magnetic material, gave birth, among others, to Mott's model of metal-insulator transition (MIT) and to the variable range hopping mechanism of electrical conductivity \cite{Mott1968, Mott1956}. Many physical properties of the material were apparently explained by a relative simple model of charge ordering of Fe$^{2+}$ and Fe$^{3+}$ cations below so-called Verwey temperature $T_{V}\approx 123$~K at which the MIT together with a structural transition occur (the Verwey model of charge ordering \cite{Verwey1939, Wright2002, Jeng2004}). 
And although in recent studies \cite{Subias2004, Shchennikov2009, Senn2012, Senn2012b} the question of the strict meaning of Verwey-like charge ordering and the low-T atomic structure is solved, the exact mechanism  of the Verwey transition is still in dispute and the phenomena below $T=50$~K are far from being well understood. These phenomena are treated in our paper. 

Magnetite is ferrimagnetic with a N\'eel temperature $T_{N} \approx 860$~K; it forms a spinel-like cubic lattice at high temperatures ($\approx 1400$~K) which gradually transforms from random spinel to inverse-spinel cubic at room temperature \cite{Mack2000}
and discontinuously transforms to monoclinic Cc at $T_V$
, as was determined by x-ray diffraction \cite{Zuo1990,Senn2012} and NMR \cite{Novak2000b}. Other measurements, however, suggest even lower crystal symmetry (magnetoelectric measurements \cite{Rado1975}, appearance of ferroelectricity \cite{Kato1981} and x-ray topography measurements \cite{Medrano1999}).

In this paper we focus on the behavior of magnetite well below $T_{V}$, especially its magnetic properties in weak ac and dc fields. We have noticed a striking difference between the ac susceptibilities accompanying magnetic after-effect studies (MAE, disaccommodation effect) of stoichiometric magnetite \cite{Walz2005} and the susceptibilities measured by other techniques (e.g. an inductive bridge method \cite{Simsa1985}) below 35~K. More specifically, the notion of the anomaly stems from the bifurcation of the $\chi'(T)$ dependence at low temperatures (see inset of FIG.~\ref{fig_anomaly}), where the upper curve represents a MAE measurement and the lower represents a classical result (e.g. a SQUID measurement; a notable exception is the result \cite{Balanda2005} of ac susceptibility studies).
Bearing in mind the specifics of MAE measurements we devised a series of experiments with the aim of explaining these discrepancies.

\section{Samples}

The samples used in our experiments were high quality single crystals of synthetic magnetite grown in the laboratory of J.M. Honig of Purdue University (samples denoted H) and by V.A.M. Brabers of Eindhoven University of Technology (samples denoted B). The samples B were grown by a floating zone technique \cite{Brabers1971}. The samples H were prepared by a skull melter technique \cite{Harrison1978}.

Namely, the sample B1 was annealed to a stoichiometric composition and cut to the shape of a cylinder 5~mm in diameter and 4~mm high. The cylinder axis is parallel to the crystallographic $c$ axis, direction $\langle 001 \rangle$. This sample was also used for NMR measurements \cite{Novak2000}. 

The sample H1 was a prism with dimensions $2.1\times2.3\times3.1\ \rm{mm}^3$ (the longest edge along $\langle 001 \rangle$) annealed for stoichiometry \cite{Aragon1983}. Since the annealing procedure was different from that for B1, it resulted in different defect pattern, as measured by MAE technique \cite{Walz2002}. This sample was also used for ac susceptibility measurements \cite{Balanda2005}. The sample H2 - a platelet of irregular shape roughly $8\times13\ \rm{mm}^2$ and thickness of 1.8~mm - although not annealed, was apparently close to stoichiometry due to preparation conditions \cite{Harrison1978}. Thus, three samples, very close to stoichiometry, but with possibly different defect pattern were measured. For comparison, the measurements were also performed on the deliberately nonstoichiometric sample, H3 a roughly pyramidal single crystal of Fe$_{3(1-\delta)}$O$_4$ with $\delta=0.002$ and dimensions approximately $1.5\times1.5\times2.0$~mm$^3$.

\section{Technique}

The magnetic properties were measured using a non-commercial, continuously operating SQUID magnetometer with an immobile sample, where the magnetic field is generated by a superconducting solenoid operating in non-persistent mode, and the variations of the magnetic moment of the sample is read continuously using a SQUID \cite{Youssef2009}. The workspace, including the superconducting solenoid, superconducting gradiometer, SQUID, thermometers, heater, and the sample was magnetically shielded, using both a superconducting (lead) can and soft magnetic material surrounding the cryostat. The achieved spectral density of magnetic moment was $\approx$17~pA~m$^2$~Hz$^{-1/2}$ and residual field was less than 2 $\mu$T inside the sample chamber. The sample, mounted on a sapphire holder, was in $^{4}$He gas environment at atmospheric pressure and was subjected to time-varying magnetic field of maximum intensity of about 20~mT. The spectral response of the magnetometer was flat from dc up to about 150~Hz.

Both the applied field generation and the sample response acquisition were carried out by a dedicated computer equipped with appropriate analog converters. In a typical susceptibility measurements reported here the applied field was varied sinusoidally, the SQUID signal (sampled at 6.4~kHz) was processed by fast Fourier transform and the results were expressed as a set of complex harmonic susceptibilities. With weak driving ac fields the sample response was linear and only the fundamental susceptibility was considered, the higher harmonics being negligible.

\section{Results}

\begin{figure}
\includegraphics [clip=true,scale=1.0]{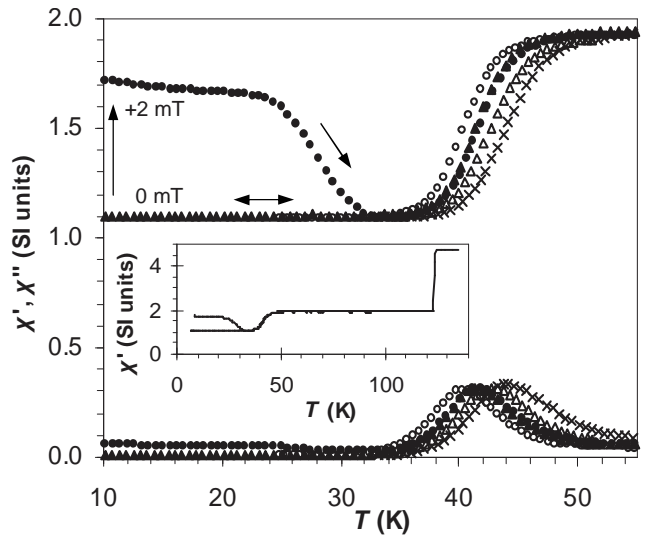} 
\caption{\label{fig_anomaly} Real part (top symbols) and imaginary part (bottom symbols) of complex ac susceptibility, H1 sample, as a function of temperature, ac field amplitude 5~$\mu$T, ac field frequency 1.6~Hz ($\circ$), 3.1~Hz ($\bullet$,$\blacktriangle$), 6.3~Hz ($\triangle$) and 12.5~Hz ($\times$).The anomaly ($\bullet$) was excited by increasing the dc field by 2~mT at 10~K. The inset: the overall temperature dependence of real susceptibility in weak magnetic field, $\mu_0 H_{ac} = 10\ \mu$T at 6.3~Hz.}
\end{figure}

When a single crystal of stoichiometric magnetite is cooled from room temperature down to a few K in the absence of a dc magnetic field and in a weak ac field, the thermal dependence of its complex ac susceptibility $\chi_{ac} = \chi' - i\chi''$ is remarkably simple (see inset of FIG. \ref{fig_anomaly}).
Above the Verwey transition the value of $\chi'$ may reach as high \cite{Simsa1985} as $10^3$ (SI units), especially near the easy axis reorientation temperature \cite{Reznicek2012}, but due to finite sample size the measured value is limited to $1/N$, $N$ being the demagnetising factor. Below $T_V$ the crystal symmetry changes and new (and large) components of magnetocrystalline anisotropy energy appear \cite{Chiba1983}, $|\chi_{ac}|$ reduces to the order of unity \cite{Simsa1985}, and demagnetising effects can be neglected in a qualitative approach.
Further below $T_V$, in the range 60~K to 35~K, a frequency-dependent thermally activated process, denoted here as "glass-like transition" after \cite{Janu2007}, occurs. Both the strong frequency dependence of this transition and the accompanying peak in absorption (the imaginary part, $\chi''$) agree well with Debye's relaxation model of thermally activated magnetic moments. Below about 30~K, when these moments are frozen in for all practical frequencies, we measured and described a novel manifestation of relaxation processes in magnetite, which we call the \emph{low temperature anomaly} (FIG. \ref{fig_anomaly}).

The anomaly is described as follows:
During cooling of a stoichiometric magnetite in a "normal" way (i.e. in relatively weak magnetic field without abrupt changes - ac field up to 100~$\mu$T and dc field up to several mT), $\chi'$ reaches its lowest value below the glass-like transition (further denoted as the "frozen" state) and retains this value down to the lowest temperatures. Also, $\chi''$ is nearly zero in this region. Upon subsequent warming at the same conditions, the same $\chi(T)$ is observed (symbols "$\blacktriangle$" in FIG. \ref{fig_anomaly}).
However, when an abrupt disturbance is applied at low temperature (e.g. step change of dc field by several mT), $\chi'$ jumps to a new higher steady value, the anomaly is "excited". Upon subsequent warming of such excited sample, $\chi(T)$ relaxes back to the "frozen" state value before the glass-like transition takes place (symbols "$\bullet$" in FIG. \ref{fig_anomaly}). This universal behaviour of stoichiometric magnetite was found to be insensitive to the ac field frequency and amplitude (provided it is relatively weak, up to about 100~$\mu$T) and the absolute intensity of dc magnetic field (within our accessible range of +/-~20~mT). Only the magnitude and character of excitation event (even a short pulse of several mT amplitude may excite the anomaly) determines the extent to which the anomaly is excited. 

\begin{figure}
\center
\includegraphics [clip=true,scale=1]{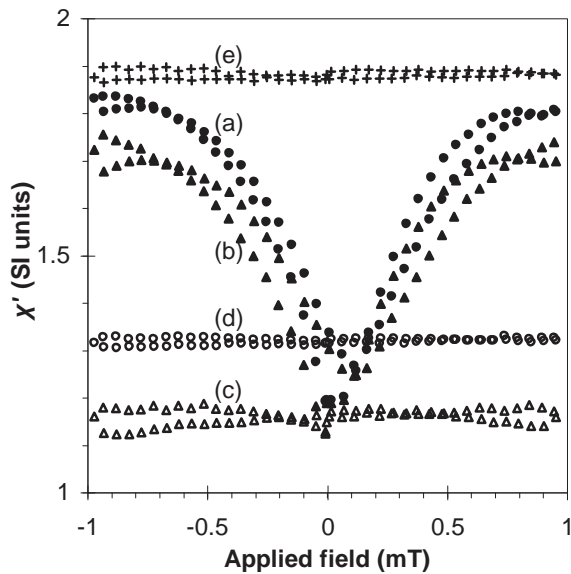} 
\caption{\label{fig_sweep} The DC field dependence of ac susceptibility at various temperatures: (a) 10~K, (b) 28~K, (c) 35~K - the "frozen" state, (d) 43~K and (e) 50~K - above the glass-like transition. The field was swept continuously in the range of +/-1~mT in a triangular fashion, the whole cycle taking 120~s. The susceptibility was measured simultaneously with an ac field of 20~$\mu$T at 6~Hz, H1 sample.}
\end{figure}

To further explore this behavior, we have performed an experiment in which the dc field was swept in a triangular fashion ( 0 $\rightarrow$ +1mT $\rightarrow$ 0 $\rightarrow$ -1mT $\rightarrow$ 0 with a period of 120~s) with superimposed weak ac field at several fixed temperatures (FIG. \ref{fig_sweep}).
Apparently, at 35~K the sample maintained the "frozen" state regardless of the changing dc field (curve (c)). At lower temperatures the change of dc field causes $\chi'$ to increase, but in this case the process is largely reversible, and the sample returns close to the "frozen" state after removal of the dc field (curves (a), (b)). With other samples (H2, B1) or with bigger changes of the dc field the process is irreversible.

\begin{figure}
\includegraphics [clip=true,scale=1]{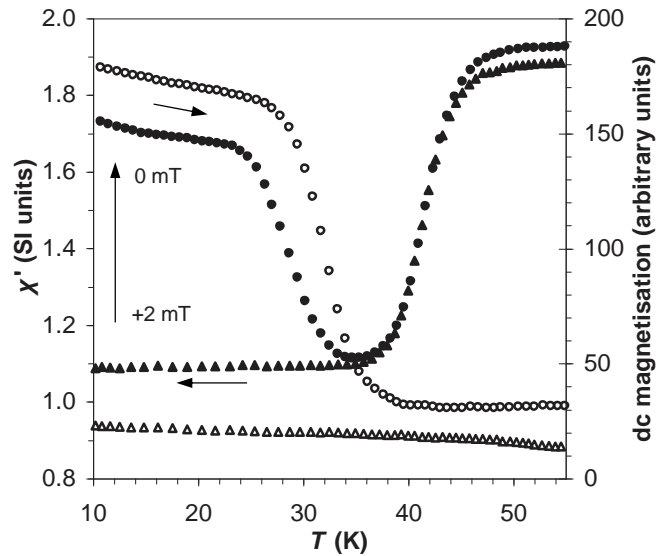} 
\caption{\label{fig_sdc} Real part of ac susceptibility (full symbols) and accompanying changes of spontaneous dc magnetisation (open symbols), H1 sample, as a function of temperature, ac field amplitude 5~$\mu$T at 6.3~Hz. The anomaly was excited by decreasing the dc field by 2~mT at 10~K. (The units of magnetisation are essentially A/m except for an additive constant - an offset - which is unknown due to working principle of a SQUID. This offset is also different for cooling ($\triangle$) and warming ($\circ$).)}
\end{figure}
Changes of spontaneous magnetisation of magnetite sample during the transition are depicted in FIG. \ref{fig_sdc}. Although more dependent on the thermal and magnetic history of the sample, the magnetisation shows several general features: (i) no dramatic changes of magnetisation occur in the region of glass-like transition, neither on cooling nor on warming the sample, (ii) if the anomaly was excited at low temperature the sample fully accommodates to the new conditions by a change of spontaneous magnetisation that takes place at higher temperature than the relaxation of ac susceptibility (to the "frozen" state), and (iii) the magnitude of the change of spontaneous magnetisation is rather low, in this case only about 10\% of the total response of the sample to the step change of the applied dc field.

\begin{figure}
\includegraphics [clip=true,scale=1]{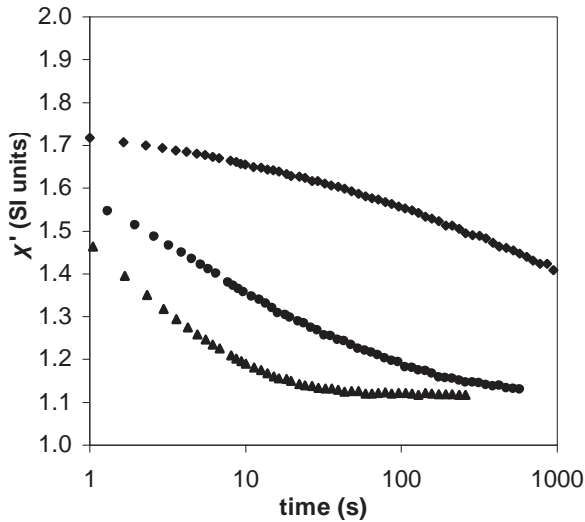} 
\caption{\label{fig_relaxation} Relaxation of the real part of ac susceptibility after a 2~mT step of applied magnetic field at 25~K (top symbols), 30~K, and 33~K (bottom symbols). H1 sample, ac field amplitude 5~$\mu$T at 6.3~Hz.}
\end{figure}

The relaxation of the anomaly was also found to depend on the rate of sample warming: the lower the rate the lower the temperature at which the "frozen" state is reached. This feature is further explored in relaxation measurements where a step change of magnetic field is applied at a defined temperature in the range of 10~K to 35~K and the time dependence of the ac susceptibility is recorded (FIG. \ref{fig_relaxation}). 

The properties of several samples (including a nonstoichiometric single crystal) were measured and the results summarised in Table \ref{table_temperatures}.
\begin{table}
\caption{\label{table_temperatures}A summary of transition temperatures (the Verwey transition, the glass-like transition $T_g$ at 6~Hz, and the temperature of relaxation of the anomaly $T_a$, warming rate 1~K/m) of measured samples. Ac amplitude was 10~$\mu$T, the anomaly was excited by a dc field step of 2~mT at 10~K. $T_g$ and $T_a$ are defined as inflection points of the $\chi'(T)$ curve. The differences in the magnitude of the anomaly (in percent of the magnitude of the glass-like transition) is mainly caused by different demagnetising factors of the samples. H3 is nonstoichiometric with $\delta = 0.002$.}
\begin{ruledtabular}
\begin {tabular}{lcccc}
sample	& $T_V$ 	& $T_g$ at 6 Hz	& $T_a$	& anomaly\\
		& (K)		& (K)			& (K) 	& magnitude			\\\hline
 H1 	& 122.2 	& 41.7 			& 27.2  & 20 \% \\
 H2 	& 122.0		& 40.3 			& 23.0	& 80 \% \\
 B1 	& 121.5 	& 36.4 			& 18.0	& 50 \% \\	
 H3 	& 117.5 	& 21.0 			& -		& - \\
\end{tabular}
\end{ruledtabular}
\end{table}
Note, that in the samples departing from perfect stoichiometry the glass-like transition shifts quickly to lower temperatures and the anomaly cannot be observed even at lowest attainable temperature (about 6~K in the current setup).

\section{Discussion}

The microscopic origin of the observed phenomena cannot be determined unambiguously from the experimental findings. However, we present several arguments suggesting that the measured changes of susceptibility of magnetite below $T_V$ arise solely from the changes of the mobility of \emph{domain walls}: (i) both the saturation magnetisation ($4\mu_B$ per Fe$_3$O$_4$ formula agrees well with the experimental value of 0.5~MA/m \cite{Ozdemir1999}) and the magnetocrystalline anisotropy (the largest component \cite{Chiba1983} is about 0.25~MJ/m$^3$) are approximately constant below $T_V$. Thus the susceptibility arising from the rotation of domain magnetisation should not depend on temperature. Moreover (ii) this susceptibility is about an order of magnitude lower \cite{Lewis1958} than the measured susceptibility. (iii) A notably similar effect was observed seven decades ago in iron with carbon or nitrogen impurities at 250~K \cite{Snoek1939}, attributed to diffusion of the impurities to the domain walls in order to accommodate magnetostrictive strain. These impurities then act as potential minima that hinder the domain wall motion in a certain temperature range.

Similar measurements of high purity single crystal magnetite \cite{Balanda2005} at low temperatures also suggest that magnetic domain walls are responsible for the observed effects. The authors also eliminated the effect of structural domain walls by detwinning the sample in strong magnetic field. However, in those measurements the anomaly always appears in the "excited" state, probably due to the high amplitude of ac magnetic field used in their experiments ($\geq$~100~$\mu$T).

To explain the low-temperature anomaly we hypothesise the existence of pinning centres that create a net of potential minima for the domain walls. Starting from the "frozen" state at 35~K (see FIG. \ref{fig_anomaly}) all domain walls lie in these minima, and a weak applied ac field induces only a small response of the sample magnetisation, $\chi'$ is small. According to the assumptions \cite{Snoek1938} about domain wall parameters we may assume that an ac field with amplitude $\mu_0 H_{ac} \leq 100\ \rm{\mu T}$ is weak enough to cause domain wall displacement that is much less than the wall thickness. This process is reversible and, if the pinning centres are stationary, no energy is dissipated ($\chi''$ is also small). But as the temperature is elevated the pinning centres are no longer fixed, thermal activation allows them to follow the displacing walls, and energy is dissipated. This process gives rise to the frequency-dependent Debye-like relaxation at 40~-~50~K as described previously \cite{Janu2007}. Above 60~K the pinning centres move freely with the domain walls, the potential minima are not effective, and $\chi'$ is large.

To complete the explanation of the anomaly let us again start from the "frozen" state at 35~K, where the domain walls reside in their potential minima. This state is preserved when the sample is further cooled to the lowest temperature (6~K in this experiment). Now the anomaly can be "excited", e.g., by changing the dc component of the applied magnetic field by a sizable amount (several mT). The domain walls are pulled from their potential minima and $\chi'$ rises (see curve (a) in FIG. \ref{fig_sweep}). When this "excited" sample is warmed gradually, the hypothetical pinning centres start to move to their new equilibrium positions, i.e. into the domain walls, new potential minima appear and the susceptibility approaches the "frozen" state. This gives rise to strongly temperature-dependent relaxation of the ac susceptibility (see FIG. \ref{fig_relaxation}), and the very same relaxation is responsible for the large peak in MAE spectra at about 30~K \cite{Walz2005}.

The connection between our results and MAE spectra is very intimate for two reasons: (i) both methods are based on ac magnetic susceptibility measurements after some excitation with a strong magnetic field (we excite the sample only once at the lowest temperature, whereas during a MAE measurement the magnetic pulse is applied periodically), and (ii) the characteristic times of both methods are similar. It is important to note that there are in fact two characteristic times incorporated in these techniques. One is determined by the frequency of probing magnetic field (1~kHz in MAE measurements and 1~Hz to 100~Hz in our measurements, corresponding to 1~ms and 10~ms to 1~s, respectively). The other characteristic time is the isochronal time of MAE experiment (1~s to 180~s); in our experiments it is determined by the rate of sample warming, and although it is not strictly defined, it is of the order of seconds to hundreds of seconds. The above mentioned peak in MAE spectra around 30~K is attributed by the authors \cite{Walz2005} to "\emph{intra-ionic} thermal inter-level excitations", i.e. an excitation of the "free" electron of an individual Fe$^{2+}$ ion from the ground sate across a gap that appears due to the energy level splitting in a trigonal crystal field. Our explanation, on the other hand, takes into account diffusion phenomena.

The microscopic nature of the hypothesised pinning centres remains unclear. However, some arguments can be summarised to elucidate the underlying mechanism: (i) a similar effect in iron \cite{Snoek1939} is ascribed to the diffusion of impurities (at concentration of 0.01\%). This most definitely will \emph{not} be the case of magnetite, as even minute concentration of impurities or structural defects shifts the glass-like transition to lower temperatures (in accordance with MAE results \cite{Walz1982}) and destroys the low-temperature anomaly (see Table \ref{table_temperatures}). Also, relaxation effects of vacancies and impurities in magnetite are expected well above $T_V$. (ii) The crystallographic domains that appear when cooling magnetite single crystal through $T_V$ have a substantial impact on the magnetic susceptibility \citep{Balanda2005}, but their structure is fixed at low temperature (easy axis switching can be induced by much larger fields close below $T_V$ \cite{Krol2007}) and the structural domains themselves cannot account for the observed phenomena. (iii) One plausible explanation is the interplay between the structural, magnetic and ferroelectric domain structure, as spontaneous polarisation was identified in magnetite thin films below 40~K \cite{Alexe2009} and the coupling between magnetic and dielectric properties of magnetite was identified long ago \cite{Rado1975}. 
%
However, a comparison of relative sizes of structural, magnetic end ferroelectric domains, that might support this explanation, cannot be easily done, as the properties of ferroelectric domains in bulk magnetite are currently unknown and the size of magnetic domains was reported in the range 30~$\mu$m to 100~$\mu$m \cite{Blackman1963,Medrano1999}. In thin samples \cite{Kasama2010} the size of both structural and magnetic domains is in the range 1~$\mu$m to 10~$\mu$m with possibly even smaller twin domains.

Other experimental findings that may be important with respect to the low-temperature anomaly are
anomalous behaviour of dielectric susceptibility \citep{Akishige1985} and significantly shortened NMR spin-spin relaxation times \cite{Novak2000},
both observed in magnetite at similar temperatures.

\section{Conclusion}

We have measured and described a novel manifestation of relaxation phenomena in high quality single crystals of stoichiometric synthetic magnetite in weak low-frequency magnetic fields. The largely unnoticed discrepancy between ac susceptibilities accompanying magnetic after-effect measurements and the results of conventional experiments (performed on top quality synthetic samples) was consistently explained as resulting from the interaction between magnetic domain walls and pinning centres that can diffuse towards the domain walls at elevated temperatures (above about 25~K).
A single diffusion process thus manifests itself both as a frequency-dependent glass-like transition above 35~K and time dependent susceptibility relaxation below 35~K, the characteristic experiment timescales being the frequency of the ac magnetic field and the rate of sample warming, respectively.

However, the microscopic origin of the pinning centres remains unclear. Impurities and structural defects that can give rise to similar efects in pure iron \cite{Snoek1939} are ruled out, as even minute amounts of defects shift the glass-like transition quickly to lower temperatures and the anomaly disappears completely. The rigidity of structural domain walls at these temperatures and the results in strong fields \cite{Balanda2005} make it evident that the interaction between structural and magnetic domains cannot account for the observed effects. A process encompassing ferroelectricity is proposed, whose elucidation will require more profound experiments to be carried out, such as combined magnetic and transport measurements in strong electric field, or nonlinear susceptibility measurements by means of higher order harmonic susceptibilities.

\begin{acknowledgments}
We would like to thank to Dr. V. A. M. Brabers of Eindhoven University of Technology, The Netherlands, for supplying us with one of the samples used in the study.

This work was supported by Institutional Research Plan Contract No. AVOZ10100520, Research Project MSM Contract No. 0021620834 and grant ESO MNT-ERA (ME10069 of MEYS CR) and SVV-2012-265303.
\end{acknowledgments} 

\bibliography{MagnetiteAnomaly}

\end{document}